# EFFECT OF ELECTROPOLISHING ON NITROGEN DOPED AND UN-DOPED NIOBIUM SURFACES*

V. Chouhan†, F. Furuta, M. Martinello, T. Ring, G. Wu, Fermi National Accelerator Laboratory, Batavia, USA

*Abstract*

Cold electropolishing (EP) of a nitrogen-doped (N-doped) niobium (Nb) superconducting RF (SRF) cavity was found to improve its quality factor. In order to understand the effect of EP temperature on N-doped and undoped surfaces, a systematic EP study was conducted with 2/0 N-doped and heat-treated Nb samples in a beaker. The Nb samples were electropolished at different surface temperatures ranging from 0 to 42 °C. The results showed that the doped surface was susceptible to the sample temperature during EP. EP resulted in the surface pitting on the doped samples where the number density of pits increased at a higher temperature. The surface results were compared with the surface of cutouts from a 9-cell cavity which was 2/0 N-doped and electropolished. This paper shows detailed surface features of the N-doped and undoped Nb surfaces electropolished at different temperatures.

## INTRODUCTION

The superconducting RF performance of niobium (Nb) cavities was greatly improved in the last 20 years. N-doping of cavities was a great invention that showed enhancement in the quality factor of the cavities [1]. This doping process has been successfully applied to LCLS II and LCLS II HE. The interstitial nitrogen atoms with no niobium nitride phase are necessary to reduce BCS resistance that enhances the quality factor [2]. After the doping is applied, niobium nitride phase formed on the surface is removed by electropolishing (EP) process. The cavities tested in a vertical cryostat revealed that the post-doping EP temperature affects their SRF performance. The cold-EP with a typical cavity temperature of ~15 ºC is applied to improve the cavity performance. This temperature was set based on the experience with the cavity performance. Although the effect of cold-EP on the cavity performance is known, a clear understanding on relation between surface feature on N-doped Nb material and a temperature in EP is still lacking. This study was conducted to gain more understanding on this subject so that further improvement on the cavity surface and performance can be achieved.

## EXPERIMENTS

EP experiments were conducted with Nb samples which were doped with the 2/0 nitrogen doping (N-doping) recipe [3] or heat-treated at 800 ºC for 3 h in a furnace. The samples during doping and heat-treatment were not covered by a Nb box or foil. The samples were experiencing furnace atmosphere during the process. The samples used in this study were prepared in a size of 1x1 cm$^2$ and from the same Nb sheet. The samples after furnace treatment were electropolished in an acid bath having a heat-sink element that maintained the acid at a desired temperature. The standard electrolyte (a mixture of sulfuric and hydrofluoric acids) was used for EP of the samples. With the setup used in this study, the acid temperature can be maintained in a wide range from ultra-cold temperature (-8 ºC or below) to room temperature. A thermocouple was attached to the Nb sample to measure its temperature during EP. This makes easy to compare sample temperature with the cavity temperature measured at the outer surface of the cavity during EP. In the sample EP, no acid stirring was applied to avoid any effect of acid flow on the surface. Sample EP was performed at different temperatures at the standard EP voltage of 18 V. A list of samples with their furnace treatment conditions and temperature in EP is given in Table 1. A removal thickness in EP for both doped and undoped samples was ~5 μm. All the samples in the EP process were set vertically with a known orientation. The sample surface was inspected with a confocal laser microscope to compare their surface features after EP.

To compare the sample surface with a cavity surface, cutouts from a 9-cell cavity (CAV0018), which was one of the early cavities treated at FNAL with 2/0 N-doping followed by EP for 7 μm average removal, were inspected with the microscope. The post-doping EP was performed at 14 V and a cavity temperature of ~24 ºC. This cavity was processed before the cold-EP temperature below 15 ºC was decided.

Table 1: N-doped and undoped samples with furnace treatment conditions and surface temperatures during EP.

| Sample | Furnace Treatment | Sample Temperature in EP (ºC) |
|---|---|---|
| 2/0-ND-1 | 2/0 N-doping | 0–3 |
| 2/0-ND-2 | 2/0 N-doping | 13–16 |
| 2/0-ND-3 | 2/0 N-doping | 22–24 |
| 2/0-ND-4 | 2/0 N-doping | 38–42 |
| 800C-UD-1 | 800 ºC/3hrs | 32–36 |

___________________
* This work was supported by the United States Department of Energy, Offices of High Energy Physics and Basic Energy Sciences under contract No. DE-AC02-07CH11359 with Fermi Research Alliance.
† vchouhan@fnal.gov

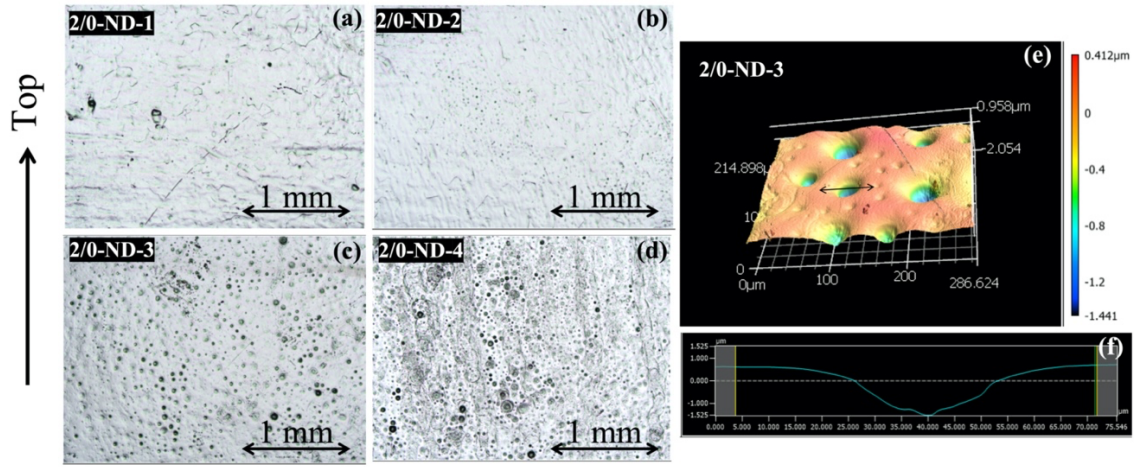

Figure 1: N-doped samples after 5 µm removal in EP performed at maximum temperatures of (a) 3 ºC, (b) 16 ºC, (c) 24 ºC, and (d) 42 ºC. The arrow shows the orientation of the samples in the EP bath. (e) 3D image of the surface shown in (c). (f) Surface profile along the line over a pit shown in (e).

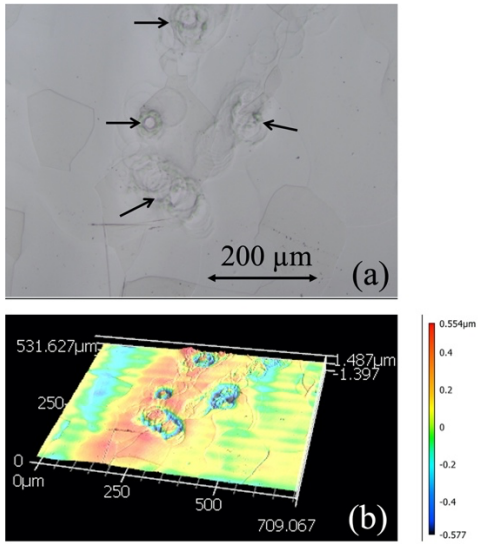

Figure 2: Traces of bubbles originating from a pit-like feature on the surface electropolished after 800 ºC/3hrs heat treatment (Sample 800C_UD-1). The arrows in (a) show the positions where the bubbles were supposed to be generated.

## RESULTS AND DISCUSSION

Microscope images of the N-doped samples listed in Table 1 are shown in Fig. 1. The orientation of the sample in the EP bath is illustrated with the images. The images showed that the surface contained many pits after removal of 5 µm. The number density of the pits was increasing with increase in the surface temperature during EP. Many pits were found after EP performed even at a low temperature ~16 ºC. However, the pits were not as wide as at higher temperatures. The depth of pits was measured to be > 2 µm as shown by a profile in Fig. 1. The surface of the samples, mainly 2/0-ND-4, electropolished at higher temperatures were also found to have traces of gas bubbles originating from the pit positions. The bubble traces were heading from the bottom to top direction of the samples set vertically during for EP. A sample was soaked in the electrolyte at 23 ºC for an extended period of 2 hours to see the effect of the acid on the surface. The surface feature changed microscopically after soaking. However, no pits were formed due to the acid soak.

The samples heat-treated (HT) at 800 ºC for 3 hours were usually not having pits except a few spots showing pit-like feature with bubble traces in a 1x1 cm$^2$ area. Figure 2 shows a spot with pit-like feature associated with bubble traces on the undoped sample (800C_UD-1). However, the pits were not as deep as observed in the case of N-doped samples. The undoped sample (UD-2), which was not subjected to any furnace treatment, have not shown any pit or bubble traces on the surface.

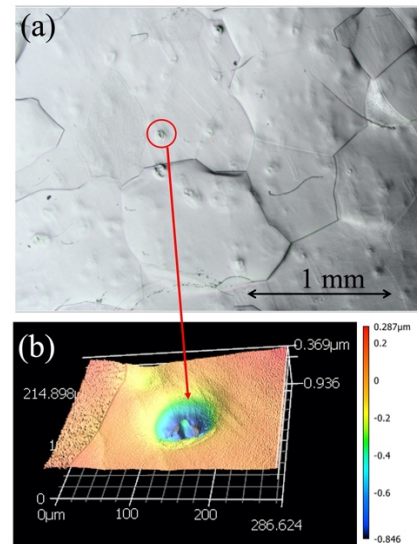

Figure 3: (a) Optical image of the cutout from cell-2 of the 9-cell cavity (b) 3D image of one of the pits on the surface.

In order to see the effect on cavities, cutout samples from a N-doped 9-cell cavity (CAV0018) electropolished for 7 µm removal were observed. EP was performed at 14 V and a cavity temperature of 24 ºC. The surface of the cutout from cell-2 of the cavity is shown in Fig. 3. The pits were also found on the cavity surface. However, the number density of pits was lower than that found on the Nb samples after EP at similar temperature. The equator surface in the end cells, which showed a rough surface due to grain etching [4], were not having pits.

The pitting on the N-doped surface was not formed after acid soaking for 2 hours. This test confirmed that the pitting was not formed due to the etching by the acid in absence of applied voltage. The pit-like feature with bubble traces on the samples suggested that gas evolution started from the spot which supposed to contain impurities. These impurities might be NbN and NbC. The generated gas bubbles were supposed to be responsible for the pits formed on the surface. The bubbles which stayed on the surface can enhance local reaction on the bubble positions [5]. The bubble traces were the signature of presence of gas bubbles.

The gas evolution from an electrode depends on an electrolyte, applied voltage, temperature, and current density. The standard acid used here is suitable for EP of Nb for a wide range of voltage, current density and temperature. The same acid may not be appropriate for EP of NbN, NbC, or other impurities in the same range of EP parameters. The sample temperature accelerates the chemical reaction rate which can elevate gas evolution rate. This might be the reason for a higher density of pits on the samples electropolished at higher temperatures.

The low number density of pits on the cavity surface as compared to that on the sample might be due to the following possible three reasons:

(1) The cavity flanges were covered with Nb caps. The caps might help to reduce incorporation of impurities into the surface [6]. The pitting and bubble traces on the 800C-UD-1 sample, which was only heat treated at 800 ºC for 3 hours without covering it with an Nb box or foil, might be due to the furnace impurity incorporated into the Nb surface.

(2) Cavity EP was performed at a lower voltage of 14 V. The low voltage EP might reduce the formation of serious pitting on the surface. The gas evolution might be also be associated with the applied EP voltage which might be too high for nitride and carbide phases of niobium to generate a gas on the surface. A typical polarization curve (a current vs voltage curve) includes a linear region, a polishing plateau region, and a gas evolution region. The gas evolution region appears at a higher voltage which varies for different materials being electropolished. Although the applied voltage was within a range of the polishing plateau for the Nb material, the same voltage might be too high to lead oxygen gas generation from the positions containing impurities. Based on this explanation, it appears that a comparatively low voltage in combination with a low temperature may reduce the number density of pits. However, reducing a voltage for EP below the plateau onset voltage results in preferential etching of Nb grains and cause a large step on the grain boundaries [5]. Etching of the grains in the end-cells of the cavity was also observed as reported elsewhere [4]. The effect of heavy doping in the end-cells and EP voltage on preferential grain etching needs to be understand. This is under study and not covered in this manuscript. Further study is necessary to understand the effect of applied voltage on the pit density and to optimize applied voltage and surface temperature for EP.

(3) The depth of pits depends on the residence time of bubble on the EP surface [5]. A longer residence time could result in a deeper pit. The cavity rotation with 1 rpm during the EP process might move the bubbles from the cavity surface and reduce the residence time of bubbles on the same position. At 1 rpm cavity rotation, the same spot on the cavity surface experienced only ~36 s in the acid since the cavity volume was ~60% filled with the acid. This obviously limited the residence time of a bubble staying on the same position to ~36 s. This suggested that the cavity rotation might be useful to reduce pitting effect.

## CONCLUSION

EP of N-doped and undoped samples was performed at different temperatures. The varying surface temperatures in the EP process affected the surface features. A higher temperature resulted in the formation of the high number density of pits which were more than 2 μm in depth. The number of pits decreased with the decrease in the sample surface temperature. Only a few pits were seen in 1x1 cm$^2$ area of the 2/0-ND-1 which was electropolished at ~0 ºC. The pits with bubble traces were found even on the electropolished sample 800C-UD-1 that was heat-treated at 800 ºC for 3 hours. This showed that furnace impurities incorporated in the Nb sample during the heat treatment. The cavity surface also showed pits which were less in the density and with smaller depth. Gas evolution from the surface was supposed to be the cause of the pit formation on the surface electropolished at higher temperatures.

A systematic study is to be performed with a coupon cavity to understand the effect of EP temperature and voltage on the N-doped cavity surface.

## ACKNOWLEDGEMENTS

This manuscript has been authored by Fermi Research Alliance, LLC under Contract No. DE-AC02-07CH11359 with the U.S. Department of Energy, Office of Science, Office of High Energy Physics.

## REFERENCES

[1] A. Grassellino *et al.*, Supercond. Sci. Technol. 26, 102001 (2013).
[2] A. Romanenko *et al.*, Appl. Phys. Lett. 105, 234103 (2014).
[3] D. Bafia *et al.*, in *Proc. of 10th Int. Part Accelerator Conf (IPAC'19)*, Melbourne, Australia, May 2019, pp. 3078–3081. doi:10.18429/JACoW-IPAC2019-WEPRB114
[4] A. Cano et al., "On the nature of surface defects found in 2/0 N-doped 9-cell cavities", presented at the 2021 Int. Conf. on RF Superconductivity (SRF'21), Michigan, USA, Jun.-Jul. 2021, paper MOPFDV009, unpublished.
[5] V. Chouhan *et al.*, *Phys. Rev. Accel. Beams,* vol. 20, p. 083502 (2017).
[6] A. Grassellino *et al*., Supercond. Sci. Technol. 30, 094004 (2017).